# A New Currency of the Future: The Novel Commodity Money with Attenuation Coefficient Based on the Logistics Cost of Anchor

Boliang Lin [a*], Ruixi Lin [b]

[a] School of Traffic and Transportation, Beijing Jiaotong University, Beijing 100044, China
[b] Department of Electrical Engineering, Stanford University, Stanford, CA 94305, USA

**Abstract:** In this paper, we reveal the attenuation mechanism of anchor of the commodity money from the perspective of logistics warehousing costs, and propose a novel Decayed Commodity Money (DCM) for the store of value across time and space. Considering the logistics cost of commodity warehousing by the third financial institution such as *London Metal Exchange*, we can award the difference between the original and the residual value of the anchor to the financial institution. This type of currency has the characteristic of self-decaying value over time. Therefore DCM has the advantages of both the commodity money which has the function of preserving wealth and credit currency without the logistics cost. In addition, DCM can also avoid the defects that precious metal money is hoarded by market and credit currency often leads to excessive liquidity. DCM is also different from virtual currency, such as bitcoin, which does not have a corresponding commodity anchor. As a conclusion, DCM can provide a new way of storing wealth for nations, corporations and individuals effectively.

## 1. Introduction

Although the commodity money has an advantage of stable intrinsic value, it has an obvious drawback of high logistics cost. In comparison, the logistics cost of metal currency is lower. To further reduce the logistics cost of commodity money, there was a time when people used precious metal as a major commodity currency. Because of the resource constraint of precious metal, the growth of precious metallic currency was far behind the accumulation of human wealth. In addition, people's expectations that precious metal will appreciate often leads to the behaviors of hoarding. This kind of phenomenon will inevitably result in liquidity shortage. Moreover, in order to reduce the quality loss and the logistics cost of metal currency in circulation, the representative money, such as silver and gold certificate, appeared. In theory, the representative money is equivalent to commodity money. In fact, the value of representative money is often lower than its anchor value represented by the goods. Obviously, the credit money or fiat money is originated from the evolution of the representative money or token money. Although the credit money has no logistics cost, it has an inherent risk of unrestrained excess issuance. Thus, it lacks the function of stable store of value. As for barter money used in barter trade, it is not easy to exchange between different countries. The virtual currencies, such as the Bitcoin, Litecoin, Primecoin, Beaocoin, Securecoin [1-5], do not create real wealth since their supply is only controlled by algorithms. Because of the fluctuation of world economy, many developed countries have started to implement the policy of negative interest rate. In this background, those credit currency holders will be the victims of inflation. Therefore, it is an important research topic on how to reveal the cost formation mechanism of symbolizing or certificating commodity money from the perspective of logistics costs, and how to design a new currency to integrate the advantages that credit money has no logistics cost and that commodity money can store wealth.

---

* Corresponding author. Email: bllin@bjtu.edu.cn

## 2. The logistics cost of symbolizing commodity money

An ideal currency should be a general equivalent of labors at a certain point of time and space, and it should have the function of store of value. Unfortunately, some credit or fiat money does not have this function.

### 2.1 The black hole of the logistics cost of banknote with anchor goods

Let us assume that an financial institution $L$ has collected material $m$ with a weight of $W_i^m$ from depositor or in general client $i$, such as 100 grams of gold, and issue a certificate or banknote $W_L^m$ to the depositor $i$ in return. In theory, the depositor can exchange certificate $W_L^m$ for any commodity in the market. The depositor may also take the banknote in return for the same amount of gold from issuer $L$. If we strictly correspond commodity $W_i^m$ with certificate $W_L^m$, the depositor can exchange certificate $W_L^m$ for commodity $W_i^m$ at any time. If so, how can we compensate issuer $L$ for its management cost? Obviously, there is a black hole of the logistics cost of anchor. One approach to fill the hole is by letting the issuer to misappropriate the depositor's goods and loan them with high interests. This approach was widely adopted by money shops in history. However, depositor will take the risk of capital loss, and it is not a good way to reserve wealth for depositors. In order to encourage financial institutions to safekeeping the goods for depositors faithfully, the depositors should pay safekeeping fees to the issuer.

### 2.2 The attenuation coefficient of the anchor of the commodity money

Generally, the logistics cost consists of ordering cost, purchase cost, transportation cost, warehouse cost, capital occupation cost etc, which can be formulated as follows[6]:

$$C^{tot}(Q) = \frac{AD}{Q} + PD + \frac{Q}{2}[S + (P+T)\lambda^{BankRate}] + TD + \lambda^{BankRate} PD \frac{t}{365} \quad (1)$$

Where, $A$ is ordering cost, $\lambda^{BankRate}$ is annual interest rate, $P$ is purchase price, $D$ is annual demand, $Q$ is the amount of each ordering, $S$ is the unit warehouse cost, $t$ is transportation time (day), $T$ is transportation cost for a product.

For the convenience of clarifying our thought, let $P_{mL}^{CIF}$ denote the price which the unit material $m$ is carried to the warehouse located at $L$, i.e., equivalent to CIF price. We have the expression as follows:

$$P_{mL}^{CIF} = \frac{A}{Q} + P + T + \frac{t\lambda^{BankRate} P}{365} \quad (2)$$

According to the formulation above, it is easy to know that the CIF price consists of contributory value of ordering cost, purchase price, transportation cost, and capital interest of materials in transport. Note that the transit time of goods is related to transportation mode. Although air transport is much faster than water, savings of time means higher transportation cost. Therefore, there is a tradeoff between the transportation cost and transit time for a company.

Obviously, the storage costs are not always the same in different sites or the same sites with different facilities. Similarly, different commodities often have different storage fees, even if they are stored in the same warehouse. Therefore the storage charges are related with community $m$ and warehouse $L$. Let $C_{mL}^{\text{warehouse}}$ denote the daily unit storage charge of $m$ in $L$. In this context, after a period of time $\Delta t$, two costs should be taken into consideration when a unit weight (or unit volume) of material $m$ is out for delivery. One is the warehouse cost as follows.

$$Z_1 = C_{mL}^{\text{warehouse}} \Delta t \tag{3}$$

The second cost is the interest of capital occupied by material.

$$Z_2 = \frac{\Delta t}{365} P_{mL}^{\text{CIF}} \lambda^{\text{BankRate}} \tag{4}$$

Hence, after a period of time $\Delta t$, the price of $m$ in $L$ should be raised to:

$$P_{mL}^{\Delta t} = P_{mL}^{\text{CIF}} + Z_1 + Z_2 \tag{5}$$

If the material needs to be delivered at a designated place, transport cost and delivery charge should be taken into account. All of these additional costs are denoted as $C_{mL}^{\text{Trans}}$. In this way, formulation (5) should be modified to be:

$$\begin{aligned} P_{mL}^{\Delta t} &= P_{mL}^{\text{CIF}} + Z_1 + Z_2 + C_{mL}^{\text{Trans}} \\ &= P_{mL}^{\text{CIF}} + C_{mL}^{\text{warehouse}} \Delta t + \frac{\Delta t}{365} P_{mL}^{\text{CIF}} \lambda^{\text{BankRate}} + C_{mL}^{\text{Trans}} \end{aligned} \tag{6}$$

After a period of time, the newly increased cost of material $m$ is:

$$P_{mL}^{\Delta t} - P_{mL}^{\text{CIF}} = C_{mL}^{\text{warehouse}} \Delta t + \frac{\Delta t}{365} P_{mL}^{\text{CIF}} \lambda^{\text{BankRate}} + C_{mL}^{\text{Trans}} \tag{7}$$

If we regard the material in storage as one of monetary anchor, we can issue representative money. For example, when the material is silver, we can issue silver certificate. The warehouse cost of this material should be reflected by the reduction of quantity marked on the certificate. To illustrate the transfer theory of warehouse cost, let $\theta_m$ denote the attenuation coefficient of anchor, where $\theta_m \in (0,1)$, and let $W_m$ be the money denomination, i.e., marked quantity of the material $m$, for instance, 100 tons of steel or 100 grams of silver. After a period of time $\Delta t$, reduction of the quantity of the material should equal to the warehouse cost and delivery charge of the third financial institution, as is shown below:

$$P_{mL}^{\text{CIF}}(1 - \theta_m^{\Delta t}) = P_{mL}^{\Delta t} - P_{mL}^{\text{CIF}} = C_{mL}^{\text{warehouse}} \Delta t + \frac{\Delta t}{365} P_{mL}^{\text{CIF}} \lambda^{\text{BankRate}} + C_{mL}^{\text{Trans}} \tag{8}$$

To obtain the attenuation coefficient, let $\Delta t = 1$, we have the formulation as follows:

$$\theta_m = 1 + \frac{\lambda^{\text{BankRate}}}{365} - \frac{C_{mL}^{\text{warehouse}} + C_{mL}^{\text{Trans}}}{P_{mL}^{\text{CIF}}} \tag{9}$$

In other words, if the financial institution sells the material $m$ with weight $W_m$ to client and keeps it for the client, the weight of this material, after a period of time $\Delta t$, will decrease to residual value:

$$W_m(\Delta t) = W_m \theta_m^{\Delta t} \quad (10)$$

In fact, the conclusion mentioned above is obtained under the assumption that the price of material $m$ is relatively stable in a time period. Otherwise, we have to consider the price fluctuation and add a new term to formulation (5). Although the material is still stored at the supplier's warehouse after the transaction, the ownership of it has shifted to the client. Thus, the capital interest is afforded by the client. Moreover, if the transport cost and delivery charge are paid on top of the attenuation coefficient, the attenuation coefficient should be modified to be:

$$\theta_m = 1 - C_{mL}^{\text{warehouse}} / P_{mL}^{\text{CIF}} \quad (11)$$

## 3. Implementation method of decayed commodity money

For some materials with stable physical property, such as base metal or precious metal, financial institution can take a single material or a combination of materials as monetary anchor. To compensate for the warehouse cost, the corresponding commodity certificate should be marked with material name, weight, grade of purity, issuance date, attenuation coefficient, delivery rules, the name of issuer, code and etc. We assume that issuer $L$ takes the material $m$ as the monetary anchor of commodity certificate, the year of issuance is $Y$ (the default date is the first day of the year), $W$ is the weight, $\theta$ denotes the attenuation coefficient, the grade of purity is denoted by $\gamma$. We define the commodity certificate with attenuation coefficient as Decayed Commodity Money (DCM). If a transaction is committed on $t_1$ days after issuance, the transaction price of this certificate should be:

$$P_{mL}^{\text{Sell}}(Y + t_1) = [\dddot{P}_m(Y + t_1) + \Delta P] W \theta^{t_1} \quad (12)$$

Where $\dddot{P}_m(Y + t_1)$ is the quotation on the international market of the material $m$ at time $Y + t_1$ corresponding to the credit money of a region, $\Delta P$ is the adjusted premium where supplier $L$ considers the purchase and transport cost. If the transaction is conducted on $t_2$ days after issuance, the actual quantity for delivery will be smaller than its marked weight $W$ and it can be calculated as follows:

$$W_{mL}^{\text{Pay}}(Y + t_2) = W \theta^{t_2} - \Delta W \quad (13)$$

The first term of the formulation above is the residual of marked weight after decay for time period $t_2$. And the second term is the additional weight reduction corresponding to delivery charge etc. Therefore DCM can be used as the reserve currency for store of value for countries, enterprises, and the public, and also as indirectly anchor for credit money or fiat money.

For example, people can pay with the gold, silver and copper certificate issued by Commodity

Exchange of New York (Comex) instead of U.S. dollar. Besides, Brent oil can be treated as the monetary anchor of DCM. Obviously, this kind of reserve currency is a good means of preserving wealth. In addition, DCM will not be hoarded by people as a result of attenuation coefficient, which can avoid the risk of liquidity shortage.

The purchase or sell of DCM can be operated in electronic accounting mode which is similar to the stock trading system, or in the account transfer mechanism from the open book management of bitcoin, or in paper money mode to circulate. Consider a part of anonymous paper bill will be lost in some case such as fire hazards, earthquakes, washing clothes. In this case, issuers will derive extra benefits. To prevent the violation of rules by financial institutions, the reserved material should be supervised by third parties.

## 4. The case study

Assume that the supplier is London Metal Exchange (LME), which is the largest nonferrous metal exchange in the world. The commodities for trading include copper, aluminum, lead, zinc, nickel, aluminum alloy. The commodity price and inventory of LME has a great influence on the production and sales of nonferrous metal worldwide.

We take copper as the monetary anchor of DCM. The copper production across the world has reached 18.43 million tons in 2014. Assume that LME takes 500 thousand tons of copper as the monetary anchor of DCM, and issues LME copper certificate on January 1st, 2020. The face value of these certificates includes 1kg, 10kg, 100kg and 1000kg. The daily attenuation coefficient is $\theta$ =99.996%, and the delivery charge is 3‰ of the actual delivery weight. The minimum delivery weight is 1000kg, and the delivery location is specified by LME. If a customer buys a LME copper certificate with a face value of 1000kg on July 1st, 2020, the residual weight after 183 days will be:

$$W(183) = 1000 \times 0.99996^{183} = 992.7066 \text{ (kg)}$$

If the copper price of the day is 5000 dollars per ton, the customer should pay 4963.5331 dollars. If the client wants to make a delivery on January 1st, 2021, the actual weight of this delivery will be:

$$W(365) = 1000 \times 0.99996^{365} \times 0.997 = 982.5493 \text{ (kg)}$$

If the customer wants to withdraw (in dollars or pounds), and the withdrawal charge is 2‰ of the actual delivery weight, the buy-back weight will be:

$$W(365) = 1000 \times 0.99996^{365} \times 0.998 = 983.5348 \text{ (kg)}$$

If the copper price is 5500 dollars per ton of the day, LME should pay to customer 5409.4414 dollars.

Let us see another example of Shanghai Futures Exchange (SHFE). The commodities for trading include copper, aluminum, lead, nickel, tin, gold, silver, deformed steel bar, wire rod and hot rolled strip etc. We consider the deformed steel bar, wire rod and hot rolled strip as standard steel ingot. Assume SHFE takes 0.4 billion tons of standard steel ingot as monetary anchor of DCM and issue steel certificate with face values including 1kg, 100kg, 1000kg, 100 tons on January 1st, 2020. The daily attenuation coefficient is $\theta$ =99.9945%, and the delivery charge is about 5‰. The minimum delivery weight is 100 tons, and the delivery location is Shanghai. This certificate is valid for 50 years. If a client buys a SHFE steel certificate with a face value of 100 tons on July 1st, 2020, the actual weight after 183 days will be:

$$W(183) = 100 \times 0.999945^{183} = 98.99856 \text{ (ton)}$$

If the standard steel ingot price of SHFE of the day is 2,500 yuan/RMB/ton, the client has to pay 247.496 thousand yuan/RMB. If the client wants to make a delivery on January 1st, 2021, the actual delivery weight will be:

$$W(365) = 100 \times 0.999945^{365} \times 0.995 = 97.5224 \text{ (ton)}$$

If the customer wants to withdraw money in Chinese yuan, and the withdrawal charge is 2‰ of the actual delivery weight, the buy-back weight will be:

$$W(365) = 100 \times 0.999945^{365} \times 0.998 = 97.8164 \text{ (ton)}$$

If the standard steel ingot price is 2,600 yuan per ton of the day, SHFE should pay to customer 254.323 thousand yuan.

## 5. Using DCM to distribute wealth across time and space

It is well known that China's real estate market is overheated at present, which implies houses have investment feature. In fact, a house is not a good anchor of commodity money. From the perspective of the logistics, the logistics cost of a house is not high in first tier cities of China, such as Shanghai, Beijing and Guangzhou, since the escalating house price and rent can offset capital interest and depreciation expense at least for now.

An investor can spend two million yuan/RMB to buy a house or buy 1000 tons of rebar and keep them in a warehouse. Comparing these two investments, the logistics cost of the latter is much higher.

Suppose SHFE issues a DCM on steel ingot, we believe that a rational investor will be more willing to buy DCM with 1000 tons steel ingot instead of a house, because DCM does not have unpredictable logistics costs, and the logistics cost is explicitly reflected by the attenuation coefficient of DCM. Investors will have a clear estimation about their wealth in the future with the DCM they buy. Once the DCM is widely known by the public, buying DCM on bulk commodity with stability physical property will become a trend. If that was the case, a present policy on limiting the production capacity of steel made by Chinese government would have been an unnecessary measure.

The total volume of steel production over the world reached 1.62 billion tons in 2015, with nearly half of it produced by China. Suppose China now takes 0.4 billion tons of steel as the monetary anchor of DCM every year, which is roughly the same amount of money of 10 million workers' annual salaries. In ten years from now, 0.1 billion tons of steel in storage will be used as storage fees of SHFE. The remaining 0.3 billion tons of steel can be translated into 7 million workers' annual salaries at that time. This means that 7 million steelworkers need not have to work in factories, instead, they can enter service industry and work for the old to alleviate the problem of supporting the old. For example, suppose a worker can serve 5 old people, then 7 million workers can support 35 million old people. If we use these 3 billion tons of steel to import labors from developing countries, we may attract 10 million foreign workers. This may relieve the labor shortage in an aging Chinese society. If China spends 2 trillion US dollars of foreign currency reserves on the purchases of copper, aluminum, silver and etc. from across the world, and store these materials as anchor, more labors will become available in the future.

## 6. Conclusions

Strictly speaking, commodity money itself is a kind of wealth, such as silver coins. A major drawback of physical money is the expensive logistics cost. When representative money comes into use, such as gold or silver certificate, the logistics cost reduces to nearly zero. However, this monetary system only works on the basis of free storage provided by financial institutions. In fact, we cannot assume free storage always exists. The financial institutions would inevitably appropriate anchor goods for other uses to gain profits and pay storage fees.

This paper presents the cost theory on the symbolization of physical currency from the perspective of logistics, and proposes a novel currency with attenuation coefficient based on the cost of anchor logistics. In this strategy, we withhold a part of anchor awarded to issuer of DCM. So DCM have the property whose value will be self-decay over time.

To sum up, DCM has the following advantages: (1) DCM has the advantage of commodity money whose intrinsic value is not easy to lose. (2) It provides a good way to distribute wealth across time and space. (3) DCM has nearly zero logistics cost, which is similar to credit money and electronic money. (4) Because of the limited resources attached to DCM, it is not likely to issue DCM excessively. (5) It can be used as the indirect anchor of credit or fiat money. (6) Due to the existence of the attenuation coefficient, it can avoid the actions of hoarding. (7) Because a part of the anchor is locked as management fees to pay to the issuer of DCM in the monetary system, the motives that issuer misappropriates anchor goods for other uses will be dispelled to a large degree. (8) DCM is corresponding to the raw materials necessary to human life, while the electronic currency such as bitcoins corresponds to mining work on computers based on algorithms, which does not produce real supplies needed by human beings.

Although DCM will lose its amount of wealth slowly over time, but it would not over-issue arbitrary amount of currencies like fiat money in some countries where there is no anchor and its wealth might suddenly evaporates.